\documentclass[onecolumn,showpacs]{revtex4}

\usepackage{graphicx}% Include figure files
\usepackage{dcolumn}% Align table columns on decimal point
\usepackage{bm}% bold math

\input epsf

\begin{document}

\title{\Large Modified Chaplygin Gas with Variable $G$ and $\Lambda$}

\author{\bf Ujjal Debnath\footnote{ujjaldebnath@yahoo.com} }

\affiliation{Department of Mathematics, Bengal Engineering and
Science University, Shibpur, Howrah-711 103, India. }

\date{\today}

\begin{abstract}
In this work, we have considered modified Chaplygin gas with
variable $G$ and $\Lambda$. The trivial solution describes
decelerating phase to accelerating phase of the universe. The
non-static with constant equation of state describes the
inflationary solution. For static universe, $G$ and $\Lambda$ must
be forms arbitrary and for static universe with constant equation
of state $G$ and $\Lambda$ should be constant.
\end{abstract}


\maketitle

\section{\normalsize\bf{Introduction}}

The Einstein field equation has two parameters - the gravitational
constant $G$ and the cosmological constant $\Lambda$. The
Newtonian constant of gravitation $G$ plays the role of a coupling
constant between geometry and matter in the Einstein field
equations. In an evolving Universe, it appears natural to look at
this ``constant'' as a function of time. Numerous suggestions
based on different arguments have been proposed in the past few
decades in which $G$ varies with time [1]. Dirac [2] proposed a
theory with variable $G$ motivated by the occurrence of large
numbers discovered by Weyl, Eddington and Dirac himself. Many
other extensions of Einstein's theory with time dependent $G$
have also been proposed in order to achieve a possible unification
of gravitation and elementary particle physics or to incorporate
Mach's principle in general relativity [3].\\

From the point of view of incorporating particle physics into
Einstein's theory of gravitation, the simplest approach is to
interpret the cosmological constant $\Lambda$ in terms of quantum
mechanics and the Physics of vacuum [4]. The $\Lambda$ term has
also been interpreted in terms of the Higgs scalar field [5].
Linde [6] proposed that $\Lambda$ is a function of temperature and
related it to the process of broken symmetries. Gaspirini [7] in
this regard argues that $\Lambda$ can also be interpreted as a
measure of temperature of a vacuum which should decrease like the
radiation temperature with cosmic acceleration. By considering
the conservation of the energy-momentum tensor of matter and
vacuum take together, many authors have invoked the idea of a
decreasing vacuum energy and hence a varying cosmological constant
$\Lambda$ with cosmic expansion in the frame work of Einstein's
theory.\\

$\Lambda$ as a function of time has also been considered by
several authors in various variable $G$ theories in different
contexts [8]. Investigating the distance dependence of gravity
under very general conditions, Wilkins [9] found that the gravity
field at a distance $r$ from a point mass has two components: one
varying as $r^{-2}$, the other as $r$ (Hookian field). The latter
component is identifiable with the weak field limit of the
$\Lambda$ term in Einstein's equation. His analysis allows one to
consider both the gravity fields - the Hookian field, coupled to
$\Lambda$ and the Newtonian one coupled to $G$ - on an equal
footing. With this in view, several authors [10, 11] have proposed
linking the variation of $G$ with that of $\Lambda$ in the
framework of general relativity. This approach preserves
conservation of the energy-momentum tensor of matter and leaves
the form of the Einstein field equations unchanged. Though this
approach is non-covariant, it is worth studying because it may be
a limit of some higher dimensional fully covariant theory [10,
12].\\

Recent observations of the luminosity of type Ia Supernovae
indicate [13, 14] an accelerated expansion of the Universe and
lead to the search for a new type of matter which violates the
strong energy condition i.e., $\rho+3p<0$. The matter consent
responsible for such a condition to be satisfied at a certain
stage of evolution of the universe is referred to as a {\it dark
energy}. There are different candidates to play the role of the
dark energy. The type of dark energy represented by a scalar field
is often called Quintessence. The simplest candidate for dark
energy is Cosmological Constant $\Lambda$. In particular one can
try another type of dark energy, the so-called Chaplygin gas which
obeys an  equation of state like [15] $p=-B/\rho, (B>0)$, where
$p$ and $\rho$ are respectively the pressure and energy density.
Subsequently the above equation was generalized to the form [16]
$p=-B/\rho^{n}, 0\le n \le 1$. There are some works on modified
Chapligin Gas obeying equation of state [17, 18]

\begin{equation}
p=A\rho-B/\rho^{n}, (A>0)
\end{equation}

At all stages it shows a mixture. This is described from radiation
era to $\Lambda$CDM model.\\

\section{\normalsize\bf{Einstein's Field Equations}}

We consider the homogeneous and isotropic space-time given by FRW
metric

\begin{equation}
ds^{2}=dt^{2}-a^{2}(t)\left[\frac{dr^{2}}{1-kr^{2}}+r^{2}(d\theta^{2}+sin^{2}\theta
d\phi^{2})\right]
\end{equation}

where $k~(=0,\pm 1)$ is the curvature parameter.\\

The energy-momentum tensor for perfect fluid is

\begin{equation}
T_{ij}=(\rho+p)u_{i}u_{j}+pg_{ij}
\end{equation}

where $\rho$ and $p$ are energy density and isotropic pressure
respectively and $c=1$.\\

The Einstein field equations with variable $G$ and $\Lambda$ is
given  by

\begin{equation}
R_{ij}-\frac{1}{2}~Rg_{ij}-\Lambda(t)g_{ij}=-8\pi G(t)T_{ij}
\end{equation}
i.e., we have two equations as

\begin{equation}
\frac{\dot{a}^{2}}{a^{2}}=\frac{8\pi G(t)
\rho}{3}+\frac{\Lambda(t)}{3}-\frac{k}{a^{2}}
\end{equation}
and
\begin{equation}
\frac{\ddot{a}}{a}=-\frac{4\pi G(t)}{3}
(\rho+3p)+\frac{\Lambda(t)}{3}
\end{equation}

From (5) and (6) we have

\begin{equation}
\dot{\rho}+3\frac{\dot{a}}{a}(\rho+p)+\rho~\frac{\dot{G}}{G}+\frac{\dot{\Lambda}}{8\pi
G}=0
\end{equation}

We assume the law of conservation of energy ($T^{ij}_{:j}=0$)
giving

\begin{equation}
\dot{\rho}+3\frac{\dot{a}}{a}(\rho+p)=0
\end{equation}

From (7) and (8) we have

\begin{equation}
\dot{\Lambda}=-8\pi \dot{G}\rho
\end{equation}

This implies $\dot{G}>$ or $<0$ according as $\dot{\Lambda}<$ or $
>0$ i.e., $G$ increases or decreases according to whether
$\Lambda$
decreases or increases.\\

Now for modified Chaplygin gas (1) we obtain

\begin{equation}
\rho=\left[\frac{B}{1+A}+\frac{C}{a^{3(1+A)(1+n)}}
\right]^{\frac{1}{1+n}}
\end{equation}

Now from equation of state (1), field equation (5) and
conservation equation (8) we have

\begin{equation}
\frac{\dot{\rho}^{2}}{\rho^{3}}=9\left(1+A-\frac{B}{\rho^{n+1}}\right)^{2}\left(\frac{8\pi
G}{3}+\frac{\Lambda}{3\rho}-\frac{k}{a^{2}\rho} \right)
\end{equation}

Differentiating w.r.t~ $t$, we obtain

\begin{equation}
\dot{\rho}\left(\frac{2\ddot{\rho}}{\rho}-\frac{3\dot{\rho}^{2}}{\rho^{2}}
\right)-\frac{3B(n+1)\dot{\rho}^{3}}{\rho^{n+3}\left(1+A-\frac{B}{\rho^{n+1}}\right)}
=3\dot{\rho}\left(1+A-\frac{B}{\rho^{n+1}}\right)\left[3\left(1+A-\frac{B}{\rho^{n+1}}\right)\left(
\frac{k}{a^{2}}-\frac{\Lambda}{3}\right)-\frac{2k}{a^{2}} \right]
\end{equation}

Now we consider $\dot{\rho}\ne 0$ (the case $\dot{\rho}=0$ i.e.,
$\rho=$ constant is discussed later). Since the above differential
equation of $\rho$ is highly non-linear, so it cannot be solve
analytically. Now for simplicity of calculation, we choose

\begin{equation}
3\left(1+A-\frac{B}{\rho^{n+1}}\right)\left[3\left(1+A-\frac{B}{\rho^{n+1}}\right)\left(
\frac{k}{a^{2}}-\frac{\Lambda}{3}\right)-\frac{2k}{a^{2}}
\right]=0
\end{equation}
and
\begin{equation}
\left(\frac{2\ddot{\rho}}{\rho}-\frac{3\dot{\rho}^{2}}{\rho^{2}}
\right)-\frac{3B(n+1)\dot{\rho}^{2}}{\rho^{n+3}\left(1+A-\frac{B}{\rho^{n+1}}\right)}
=0
\end{equation}

From equation (13) we find the trivial solution of $\Lambda$ i.e.,

\begin{equation}
\Lambda=\frac{(1+3A)k}{(1+A)a^{2}}-\frac{2kB}{C(1+A)^{2}}~a^{3(1+A)(1+n)-2}
\end{equation}

Also from equation (14) we obtain (after manipulation)

\begin{equation}
a^{\frac{3(1+A)}{2}}~_{2}F_{1}[\frac{1}{2(1+A)},\frac{1}{2(1+n)},1+\frac{1}{2(1+n)},
-\frac{Ba^{3(1+A)(1+n)}}{C(1+A)}]=\frac{\rho_{0}}{2}~C^{\frac{1}{2(1+n)}}~t
\end{equation}

For $k>0$, $\Lambda$ decreases with $t$ upto certain stage of the
evolution of the universe and for $k<0$, $\Lambda$ increases with
$t$ after certain stage of the evolution of the universe (see
figure 1) and
for $k=0$, we must have $\Lambda=0$.\\

From (9) and (15) we have

\begin{equation}
4\pi
G=\frac{\frac{(1+3A)k}{(1+A)a^{3}}+\frac{kB\{3(1+A)(1+n)-2\}}{C(1+A)^{2}}~a^{3\{(1+A)(1+n)-1\}}}
{\left[\frac{B}{1+A}+\frac{C}{a^{3(1+A)(1+n)}}
\right]^{\frac{1}{1+n}}}
\end{equation}

Note that here $G_{0}$ depends on the value of $k$ i.e., $k=0$ implies $G_{0}=0$. From figure 2
we see that $G$ increases with the evolution of the universe for $k=1$.\\

For early universe i.e., for $a\approx 0$, we get

\begin{equation}
4\pi G \approx
\frac{(1+3A)k~a^{1+3A}}{(1+A)C^{\frac{1}{1+n}}}+4\pi
G_{0}~~,~~~~~~(G_{0}>0)
\end{equation}

i.e., $G\rightarrow G_{0}$ as $a\rightarrow 0$.\\

Also for late universe i.e., for $a\approx\infty$, we have

\begin{equation}
4\pi G \approx \frac{kB}{C(1+A)^{2}}\left(\frac{1+A}{B}
\right)^{\frac{1}{1+n}}~a^{3(1+A)(1+n)-2}+4\pi G_{0}
\end{equation}

i.e., $G\rightarrow\infty$ as $a\rightarrow\infty$.\\

\begin{figure}
\includegraphics[height=2.0in]{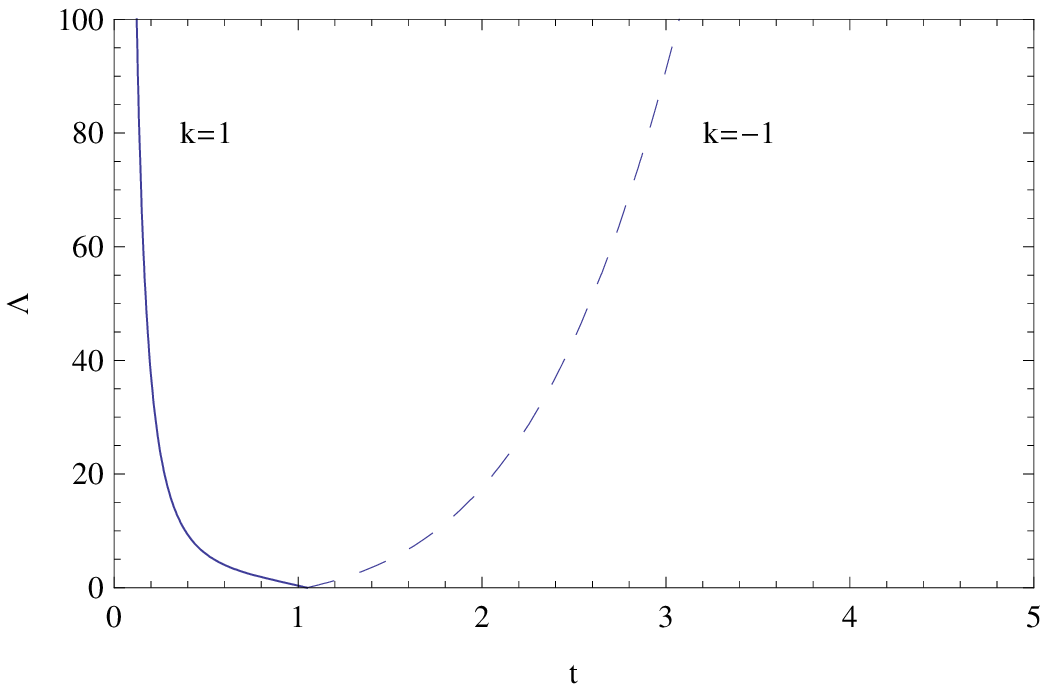}~~~
\includegraphics[height=2.0in]{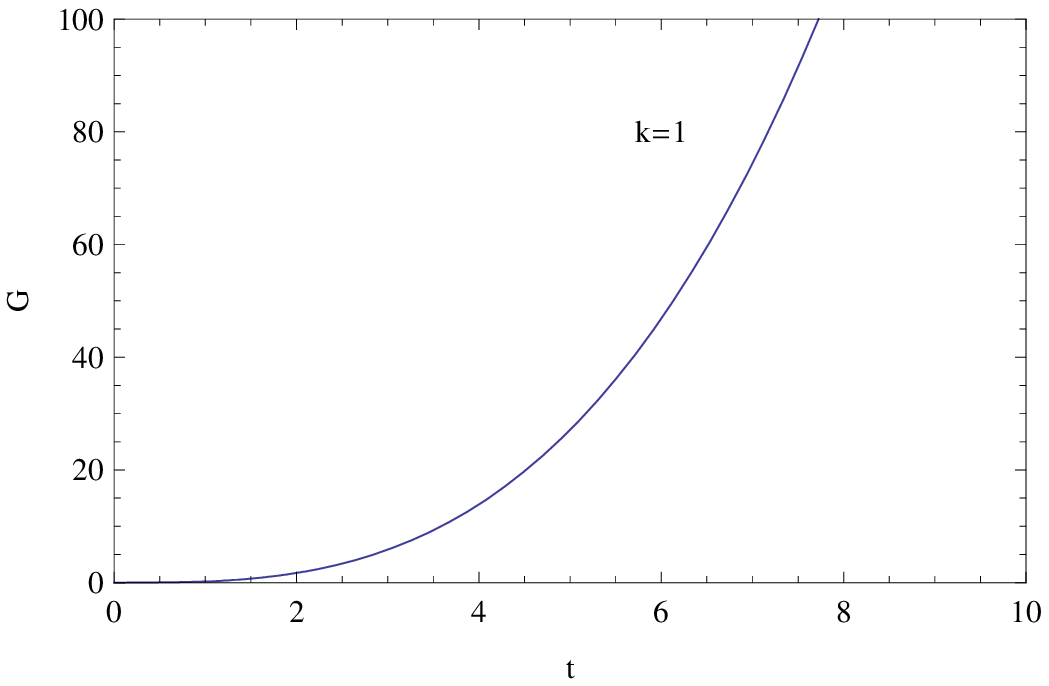}\\
\vspace{1mm}
~~~~~Fig.1~~~~~~~~~~~~~~~~~~~~~~~~~~~~~~~~~~~~~~~~~~~~~~~~~~~~~~~~~~~~~~~~~~~~~~~~~~~~~Fig.2\\

\vspace{10mm}

\includegraphics[height=2.0in]{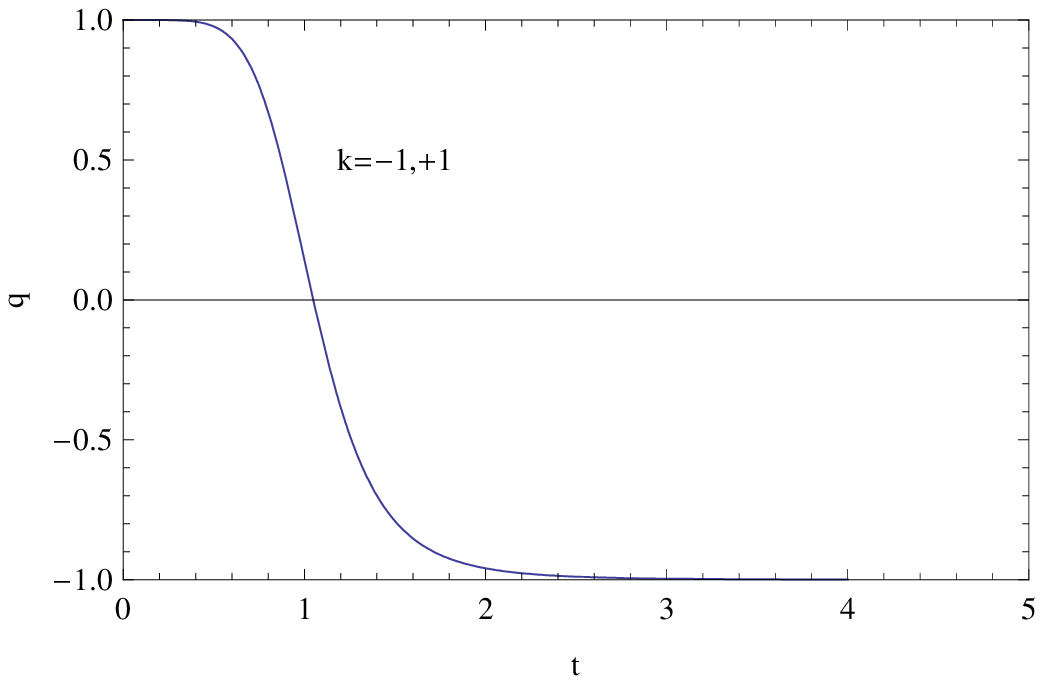}
\\
\vspace{1mm} ~~~~~Fig.3

\vspace{6mm}

Fig. 1, 2 and 3 show the variations of $\Lambda$, $G$ and $q$
against $t$ for $A=1/3,~B=1,~C=1,~n=1/2$. \vspace{4mm}

\end{figure}

The deceleration parameter has the expression

\begin{equation}
q=-\frac{a\ddot{a}}{\dot{a}^{2}}=\frac{4\pi
G(\rho+3p)-\Lambda}{8\pi G\rho+\Lambda-3ka^{-2}}
\end{equation}

From figure 3 we see that $q$ decreases from $+1$ to $-1$ for
$A=1/3,~k=\pm 1$. This implies the universe has early
deceleration and late acceleration. For $k=0$, we have
$\Lambda=0$ and $G=0$. So the field equations
(5) and (6) yield $a=$ constant, which is static solution.\\

The above discussions are valid for $\dot{\rho}\ne 0$ i.e.,
$\rho\neq$ constant. Now we will discuss in the case
$\dot{\rho}=0$ i.e., $\rho=$ constant $=\rho_{0}$ (say).\\

Now equation (8) reduces to

\begin{equation}
(\rho+p)~\frac{\dot{a}}{a}=0
\end{equation}

So we have possibilities:  $\rho+p=0$ or $\dot{a}=0$.\\

(i) $\dot{a}\neq 0$ and $\rho+p=0$: This implies
$p=-\rho=-\rho_{0}=p_{0}$ (say). From equation (1) we have
$\rho_{0}=\left(\frac{B}{1+A}\right)^{\frac{1}{1+n}}$. So from
field equations (5) and (6), we get
$$
a=\sqrt{\frac{k}{C_{1}}}~\text{cosh}(\sqrt{C_{1}}~t)
$$

From equation (9) we have

\begin{equation}
\Lambda=\Lambda_{0}-8\pi
G\left(\frac{B}{1+A}\right)^{\frac{1}{1+n}}~~,
~~~~(\Lambda_{0}=3C_{1})
\end{equation}

There are no other equations, so $\Lambda$ and $G$ can not be
calculated. We also see that in this case, $G$ increases or
decreases according as $\Lambda$ decreases or increases.
$\Lambda$ and $G$ are arbitrary functions of time in this case.\\

The field equations (5) and (6) and equation (22), we have

$$
t=\sqrt{\frac{3}{\Lambda_{0}}}~\text{log}\left(\frac{\sqrt{\Lambda_{0}}~a+
\sqrt{\Lambda_{0}a^{2}-3k}}{\sqrt{\Lambda_{0}}~a_{0}+
\sqrt{\Lambda_{0}a_{0}^{2}-3k}}\right)
$$
i.e.,
\begin{equation}
a=a_{0}~\text{cosh}(\sqrt{\frac{\Lambda_{0}}{3}}~t)+\sqrt{a_{0}^{2}-\frac{3k}{\Lambda_{0}}}~\text{sinh}(\sqrt{\frac{\Lambda_{0}}{3}}~t)
\end{equation}

For $k=0$ we have $a=a_{0}~e^{\sqrt{\frac{\Lambda_{0}}{3}}~t}$
which is the inflationary solution.\\

(ii) $\dot{a}=0$ and $\rho+p\ne 0$: This implies $a=$ constant =
$a_{0}$ i.e., we have static universe. In this case, from field
equations we have the values of $G$ and $\Lambda$ as

\begin{equation}
G=\frac{k}{4\pi
a_{0}^{2}\left[(1+A)\rho_{0}-\frac{B}{\rho_{0}^{n}}
\right]}=\text{constant}
\end{equation}
and
\begin{equation}
\Lambda=\frac{k}{a_{0}^{2}}~\left[\frac{(1+3A)\rho_{0}^{n+1}-3B}{(1+A)\rho_{0}^{n+1}-B}\right]=\text{constant}
\end{equation}

where $\rho_{0}\ne \left(\frac{B}{1+A} \right)^{\frac{1}{n+1}}$.\\

(iii) $\dot{a}=0$ and $\rho+p=0$: This implies $a=$ constant =
$a_{0}$ i.e., we also have static universe and
$p=-\rho=-\rho_{0}=p_{0}$ (say). From equation (1) we have
$\rho_{0}= \left(\frac{B}{1+A} \right)^{\frac{1}{n+1}}$. From
field equations, $G$ and $\Lambda$ satisfies:

\begin{equation}
\Lambda+8\pi G\rho_{0}=\frac{3k}{a_{0}^{2}}
\end{equation}
and
\begin{equation}
\Lambda+8\pi G\rho_{0}=0
\end{equation}

which are consistent only for $k=0$. This implies

\begin{equation}
\Lambda=-8\pi G\rho_{0}=8\pi G\left(\frac{B}{1+A}
\right)^{\frac{1}{n+1}}
\end{equation}

The above relation (28) shows that $G$ and $\Lambda$ are
arbitrary functions of time $t$.\\

\section{\normalsize\bf{Concluding Remarks}}

From the field equations and the conservation equation, an
equation is obtained in $\rho$, $a$ and $\Lambda$ which suggests
two trivial solutions. The trivial case with $\Lambda$ is given
in equation (15) leads to a model which starts from big bang with
non-zero gravitational constant $G_{0}$ ($k\ne 0$) and has a
positive constant deceleration parameter $q=1$ and leads to
accelerating universe $(q=-1)$ with infinite gravitational
constant $G$ (see figure 3). For $k=0$ we have $\Lambda=0$ and
$G=0$ and $a=$ constant, which is static solution. The model
describes the evolution of the universe with early deceleration
and late acceleration. For constant density ($\rho_{0}\ne
\left(\frac{B}{1+A} \right)^{\frac{1}{n+1}}$) universe, the
cosmological constant $\Lambda$ and gravitational constant $G$
are arbitrary functions of time in which for static universe
describes $k=0$. But for static universe with non-constant
density, $G$ and $\Lambda$ are constant where constant density
$\rho_{0}\ne \left(\frac{B}{1+A} \right)^{\frac{1}{n+1}}$.\\\\

{\bf Acknowledgement:}\\

The author thanks IUCAA for worm hospitality where the part of the
work was done.\\

{\bf References:}\\
\\
$[1]$ P. S. Wesson (1978), {\it Cosmology and Geophysics} (Oxford
: Oxford University Press ); P. S. Wesson (1980), {\it Gravity,
Particles and Astrophysics} ( Dordrecht : Rieded ).\\
$[2]$ P. A. M.  Dirac, {\it Proc. R. Soc. A} {\bf 165} 119 (1938)
; {\bf 365} 19 (1979) ; {\bf 333} 403 (1973); The General Theory
of Relativity (New York : Wiley) 1975.\\
$[3]$ F. Hoyle and J. V. Narlikar, {\it Proc. R. Soc. A} {\bf 282}
191 (1964); {\it Nature} {\bf 233} 41 (1971); C. Brans and R. H.
Dicke, {\it Phys. Rev.} {\bf 124} 925 (1961).\\
$[4]$ Y. B. Zeldovich, {\it Sov. Phys.- JETP} {\bf 14} 1143
(1968); {\it Usp. Fiz. Nauk} {\bf 11} 384 (1968); P. J. E. Peebles
and B. Ratra, {\it Astrophys. J.} {\bf 325} L17 (1988).\\
$[5]$ P. G. Bergmann, {\it Int. J. Theor. Phys.} {\bf 1} 25
(1968); R. V. Agoner, {\it Phys. Rev. D} {\bf 1} 3209 (1970).\\
$[6]$ A. D. Linde, {\it JETP Lett.} {\bf 19} 183 (1974).\\
$[7]$ M. Gasperini, {\it Phys. Lett.} {\bf 194B} 347 (1987); {\it
Class. Quantum Grav.} {\bf 5} 521 (1988).\\
$[8]$ A. Banerjee, S. B. Dutta Chaudhuri and N. Banerjee, {\it
Phys. Rev. D} {\bf 32} 3096 (1985); O. Bertolami {\it Nuovo
Cimento} {\bf 93B} 36 (1986); {\it Fortschr. Phys.} {\bf 34} 829
(1986); Abdussattar and R. G. Vishwakarma, {\it Class. Quantum
Grav.} {\bf 14} 945 (1997); Arbab I. Arbab, {\it Class. Quantum Grav.}
{\bf 20} 93 (2003).\\
$[9]$ D. Wilkins, {\it Am. J. Phys.} {\bf 54} 726 (1986).\\
$[10]$ D. Kaligas, P. Wesson and C. W. F. Everitt, {\it Gen. Rel.
Grav.} {\bf 24} 351 (1992).\\
$[11]$ A - M. M. Abdel Rahaman, {\it Gen. Rel. Grav.} {\bf 22} 655
(1990); M. S. Berman, {\it Gen. Rel. Grav.} {\bf 23} 465 (1991);
A. Beesham, {\it Int. J. Theor. Phys.} {\bf 25} 1295 (1986).\\
$[12]$ P. S. Wesson, {\it Gen. Rel. Grav.} {\bf 16} 193 (1984).\\
$[13]$ N. A. Bachall, J. P. Ostriker, S. Perlmutter and P. J.
Steinhardt, {\it Science} {\it 284} {\it 1481} (1999).\\
$[14]$ S. J. Perlmutter et al, {\it Astrophys. J.} {\bf 517} 565
(1999).\\
$[15]$ A. Kamenshchik, U. Moschella and V. Pasquier, {\it Phys.
Lett. B} {\bf 511} 265 (2001); V. Gorini, A. Kamenshchik,
U. Moschella and V. Pasquier, {\it gr-qc/0403062}.\\
$[16]$ V. Gorini, A. Kamenshchik and U. Moschella, {\it Phys. Rev.
D} {\bf 67} 063509 (2003); U. Alam, V. Sahni , T. D. Saini and
A.A. Starobinsky, {\it Mon. Not. Roy. Astron. Soc.} {\bf 344},
1057 (2003); M. C. Bento, O. Bertolami and A. A. Sen, {\it Phys.
Rev. D} {66} 043507 (2002).\\
$[17]$ H. B. Benaoum, {\it hep-th}/0205140.\\
$[18]$ U. Debnath, A. Banerjee and S. Chakraborty, {\it Class.
Quantum Grav.} {\bf 21} 5609 (2004).\\

\end{document}